\documentclass[12pt,a4paper]{article}%
\usepackage{a4wide}
\usepackage{latexsym}
\usepackage{epsf}
\usepackage{amssymb}
\usepackage{graphicx}
\usepackage{amsmath, cite}
\usepackage{amsmath,amssymb,amsthm}
\usepackage{verbatim}
\usepackage{hyperref}%
\usepackage{amsmath}%
\usepackage{amsfonts}
\renewcommand{\d}{\textrm{d}}

\begin{document}
\numberwithin{equation}{section}
\begin{flushright}
\small
UUITP-03/12
\normalsize
\end{flushright}

\begin{center}
{\LARGE \textbf{{Resolving anti-brane singularities
\\ \vspace{0.5cm} through time-dependence}}}

\vspace{1 cm} {\large Johan Bl{\aa }b\"ack$^{a}$, Ulf H.
Danielsson$^{a}$, Thomas Van Riet$^{b}$}\\

\vspace{0.85 cm}{$^{a}$ Institutionen f{\"o}r fysik och astronomi,
Uppsala Universitet,\\  Box 803, SE-751 08 Uppsala, Sweden}
\\\vspace{0.4 cm}

{$^{b}$ Institut de Physique Th\'eorique, CEA Saclay, CNRS URA 2306,
\\  F-91191 Gif-sur-Yvette, France}

\vspace{0.7cm} {\small\upshape\ttfamily johan.blaback,
ulf.danielsson @
physics.uu.se, thomas.van-riet @ cea.fr} \\

\vspace{4cm}

\textbf{Abstract}
\end{center}

\begin{quotation}
In this note we discuss a possible resolution of the flux
singularities associated with the insertion of branes in backgrounds
supported by fluxes that carry charges opposite to the branes. We
present qualitative arguments that such a setup could be unstable
both in the closed and open string sector. The singularities in the
fluxes then get naturally resolved by taking the true solution to be
a time-dependent process in which flux gets attracted towards the
brane and subsequently annihilates.

\end{quotation}

\newpage

\section{The problem setting}

It is of great importance to understand the non-supersymmetric
supergravity solutions that describe anti-D$3$ branes in the
Klebanov--Strassler (KS) throat \cite{Klebanov:2000hb} for their
importance in de Sitter model building \cite{Kachru:2003aw,
Balasubramanian:2005zx} and holographic duals to non-supersymmetric
confining gauge theories \cite{Kachru:2002gs}. Impressive progress
on this has recently been made in \cite{Bena:2009xk, Bena:2011hz,
Bena:2011wh}, improving on earlier works \cite{DeWolfe:2008zy,
McGuirk:2009xx}.

The three-form fluxes in the KS throat effectively induce D$3$
charges and this is why the insertion of anti-D$3$ branes breaks
supersymmetry and the solutions become very involved. One of the
eye-catching properties of the solutions is the presence of singular
three-form fluxes near the anti-D$3$ branes, which was first noted
in \cite{McGuirk:2009xx}. These singular fluxes are problematic
since they cannot simply be resolved in the same way as the
$F_{5}$-flux singularity. The latter flux singularity is understood
as it is part of the standard singular fields directly sourced by
the branes and in full string theory D$3$-branes are regular
objects. However it has been argued in \cite{Dymarsky:2011pm} that
the singular three-form fluxes are an artefact of the perturbation
theory employed in \cite{Bena:2009xk, Bena:2011hz, Bena:2011wh}.
Other suggestions are that the singularities are related to the
smearing of the anti-D$3$ branes over the $S^{3}$-part of the warped
conifold and that a full analysis, in the style of
\cite{Polchinski:2000uf}, would resolve the singularities
\cite{Dymarsky:2011pm}. In this paper we interpret the singularity instead as an indication of instability.

Analogues supergravity solutions that
describe anti-D$p$ branes in flux backgrounds of opposite charge
have also been considered for anti-D$2$ branes
\cite{Giecold:2011gw}, anti-M$2$ branes \cite{Bena:2010gs,
Massai:2011vi} and anti-D$6$ branes \cite{Blaback:2011nz,
Blaback:2011pn}. The way fluxes can induce magnetic D$p$ brane charges occurs through
the transgression terms inside the Bianchi identity:
\begin{equation}
\label{Bianchi}\d F_{8-p} = H \wedge F_{6-p}\,,
\end{equation}
with $p$ and integer between $0$ and $6$. According to this identity
a suitable combination of $H$- and $F_{6-p}$-flux can induce the
same charges as an D$p$ brane that magnetically sources $F_{8-p}$.
If one then adds explicit anti-D$p$ branes to this background one
obtains non-SUSY solutions since the anti-D$p$ branes carry opposite
charges with respect to the fluxes. In case the fluxes would be
replaced by D$p$-branes the insertion of the anti-branes would lead
to an unstable background in which all anti-branes annihilate. In
the case of fluxes there is still a similar annihilation, but it is
claimed to have a classical barrier against this annihilation when
the anti-brane charge is small enough compared to the background
flux \cite{Kachru:2002gs}. In this note we reconsider these claims.

All the anti-brane solutions constructed so far share a singular
blow-up of (one or both of) the fields that appear on the right hand
side of the Bianchi-identity (\ref{Bianchi}). Only for the simplest
case of $p=6$ have these singularities been established beyond
perturbation theory and with fully localised sources
\cite{Blaback:2011nz, Blaback:2011pn}\footnote{The BPS background
one obtains by replacing the SUSY-breaking anti-D$6$ branes with
D$6$ branes has been discussed in \cite{Janssen:1999sa}.} \footnote{After the first version of this paper appeared it was demonstrated a few months later, by different authors, that the anti-D3 singularity is there at all orders, at least when the anti-D3 brane is smeared over the tip of the KS cone \cite{Bena:2012bk}. }. Since all
the anti-brane solutions are similar, and roughly related by
T-duality, this can be considered as a good indication that the
singularities are present also for the (much more complicated)
anti-D$2$, anti-D$3$ and anti-$M2$ brane solutions. It is important to understand what the nature of these flux singularities is and it  is the aim of this paper to  provide a possible interpretation.  

As explained in
\cite{Blaback:2010sj, Blaback:2011nz, Blaback:2011pn} it is not
surprising that something must happen with the flux combination $H
\wedge F_{6-p}$ near the anti-D$p$ branes, because this flux is
electromagnetically and gravitationally attracted towards the
anti-brane.  For example, the force on a probe D3 in the
linearised anti-D3 solution of \cite{Bena:2009xk} gives a force towards the
anti-brane \cite{Bena:2010ze}. The flux carries the same charge and mass, and should hence
also be attracted towards the anti-brane.
Subsequently one can interpret the singularity as the
inability of the system to find a balance against this flux
attraction\footnote{The arguments presented in
\cite{Dymarsky:2011pm} that attribute the singularities to the
perturbation theory is only valid for some of the components of the
$G_3$ flux \cite{Massai:2012jn}.}. In
this note we develop this picture in more detail and we argue that
the backreaction of the anti-branes is not localised in time,
whereas it is in space. We can then formulate a consistent
time-dependent physical picture in which all fields are regular at
all times.

Finally we like to mention that recently an example has been found
in which anti-brane backreaction is not localised in space, and this
(perhaps counter-intuitive) behaviour can invalidate specific
inflationary models of large-field inflation \cite{Conlon:2011qp}.
This illustrates the constraining character of string theory when
building phenomenologically interesting models that break
supersymmetry.


\section{A simple analogy}

It is insightful to undo the problem of all its complications and
non-linearities associated with (super-)gravity and consider a
simpler problem with analogous physics. Such an analogy can be found
in electrodynamics. Consider an infinite cloud of a positively
charged fluid. The fluid has an internal electrostatic repulsion and
the cloud's lowest energy state (the SUSY vacuum) is the one in which
the fluid is uniformly distributed. The characteristic quantity of
the cloud is the charge density $\rho$.
Now we ``pollute'' this system by inserting a negatively charged
particle into it with total negative charge equal to $p$. Clearly
what will happen is that the cloud will try to screen the negative
charge by clumping around it as depicted in figure \ref{Figure1}
below. In the case of fluxes and anti-branes, such clumping of the
flux near the anti-brane has first been mentioned in
\cite{DeWolfe:2004qx}.

\begin{figure}[h]
\centering
\includegraphics[width=.4\textwidth]{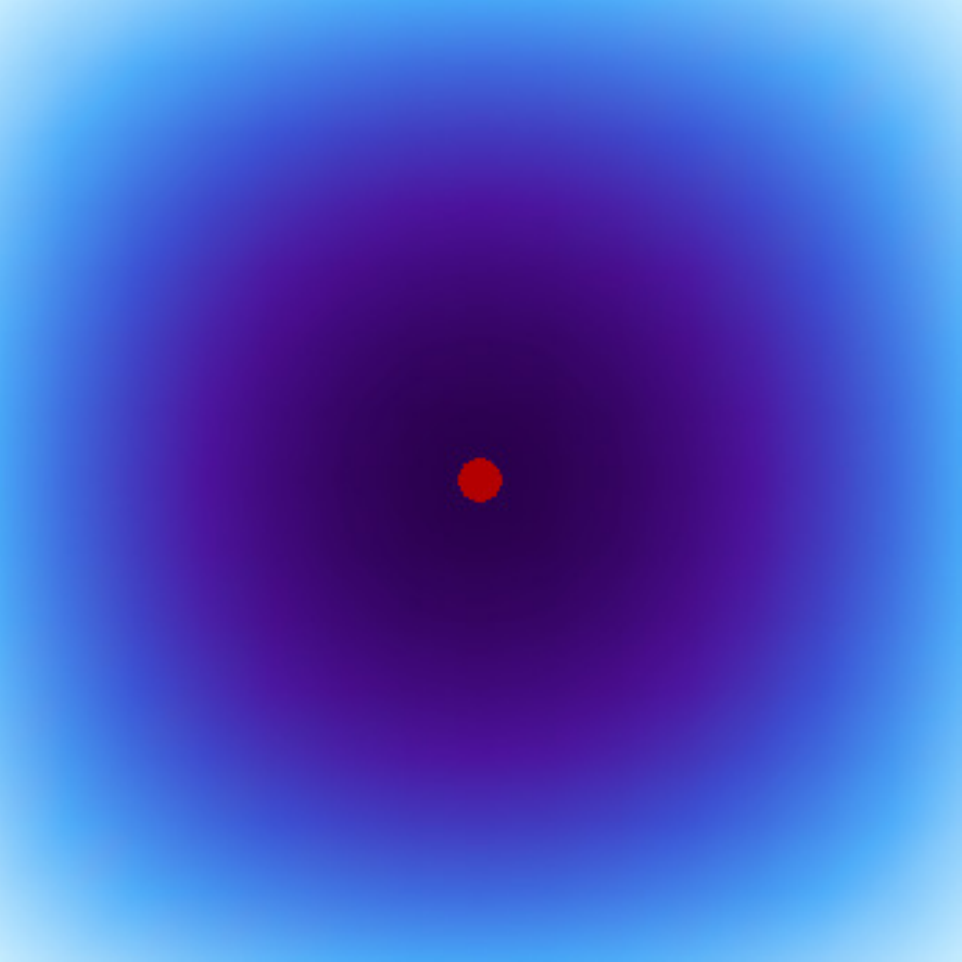}\caption{The clumping of a positively charged
fluid near a negatively charged particle. }%
\label{Figure1}%
\end{figure}

The details of the clumping depend on the specific details of the interactions
between the fluid and the particle. We envisage the following interactions, in
order to copy the brane-flux behaviour. There is an interaction term in the
full Hamiltonian that can describe the annihilation between positive and
negative charges.
For brane-flux annihilation it is known that for $p$ small enough there is a
barrier against direct annihilation between fluxes and branes. Similarly, here
we assume the same to be true. It is furthermore sensible to assume that the
annihilation process is facilitated when there is a higher cloud density
$\rho$. Hence for a fixed $p$ there will be a critical density $\rho_{c}(p)$,
such that for higher densities the barrier against direct annihilation
vanishes. Below we will verify explicitly that this is the case for brane-flux
annihilation. Secondly we assume that the cloud has some pressure, partly due
to the electrostatic repulsion. The analogy with the flux cloud in
supergravity is that flux wants to be as homogeneous as possible (for a given
flux number) in order to achieve the lowest gradient energy
\begin{equation}
E \propto\int\sqrt{|g|}F_{\mu\nu}F^{\mu\nu}\,.
\end{equation}
These two terms in the Hamiltonian, the pressure and the annihilation, leave
two possibilities for the evolution of the cloud:

\begin{enumerate}
\item If the repulsion in the cloud is large enough clumping will not be too
severe and there is a barrier against annihilation. Hence there will
a classical stationary solution reflecting the meta-stable state.
Only quantum mechanical tunneling will bring one to the final stable
(SUSY) ground state in which the negative charges are annihilated
and a positively charged cloud with a somewhat smaller charge
density $\rho$ emerges.

\item The repulsion energy is not strong enough and the system changes in time
as clumping increases $\rho$ locally. After some time the density $\rho$
around the negative charge becomes larger than the critical density $\rho>
\rho_{c}$ and the barrier against annihilation vanishes. Therefore there is no
meta-stable stationary state, rather a time-dependent solution that describes
a clumping cloud in which the screened charge annihilates and one decays
perturbatively into the SUSY ground state.
\end{enumerate}

In the first case there would exist a classical stationary solution
to the equations of motion, without the annihilation process at
work. The dynamics without annihilation is the analogy of
supergravity equations of motion (closed string interactions) and
the annihilation is the analogue of open string effects (as it
proceeds via brane nucleation). Therefore having a reasonable static
solution to the supergravity equations of motion is a prerequisite
for realizing the first scenario. We have mentioned in the
introduction that the stationary solutions that describe the
anti-D$p$ branes in a flux background that carries D$p$-charge have
divergent fluxes. This points towards the second scenario. If the
clumping of the cloud would be counterbalanced by its internal
pressure to give a stationary solution, it would do so with
\emph{finite} fluxes everywhere. As explained in
\cite{Blaback:2011nz, Blaback:2011pn} the divergent (infinite)
fluxes are such that they can be interpreted as a clumping of flux
with a charge that is trying to screen the anti-brane charge. One
should also note that it is a common property of perturbation theory
to give IR singularities when expanding around a bad background.

Another insightful analogy is that of black hole formation when the thermal pressure of a star 
cannot counterbalance the gravitational attraction anymore. The singular black hole solution is the only static solution and is not describing the real physics, it only describes the physics
at large times. The real solution is a time-dependent smooth solution that becomes singular at large times. 
Our claim is that the flux singularity should be seen as exactly this: the real solution is time-dependent since the flux moves towards the brane. At late times it settles into a singular static solution.  Below we  further claim that the open string degrees of freedom can then cure the singularity by letting the flux annihilate\footnote{Also here is a similarity with black holes.
One expects that black holes are regular in a full quantum gravity treatment, where more degrees of freedom are included. However, the analogy has its limitations since a black hole final state has all matter behind the horizon and there is no notion of thermal pressure anymore. For the singular flux solutions, we do not have that all flux collapsed to a single point. There is still flux pressure, but the flux has been piled up in an unphysical, singular, matter.}.

\section{Brane/flux annihilation}

For the case of anti-D$3$ branes in the Klebanov--Strassler (KS)
background it
has been claimed by Kachru, Pearson and Verlinde (KPV) in \cite{Kachru:2002gs}%
, that for tunably small $p$ (compared to the background flux number
$M$) there should indeed be a barrier against the annihilation
process. The computation for this is carried out using probe
NS$5$-brane actions within the undeformed KS geometry. The reason
NS$5$-branes are used is that the anti-D$3$ branes puff up into
NS$5$-branes, according to the Myers effect \cite{Myers:1999ps}.
These NS$5$-branes wrap an $S^{2}$ and have a non-zero worldvolume
flux, such that the puffed NS$5$ induces $p$ units of anti-D$3$
charge and tension. When the NS$5$-brane moves towards the other
pole of the $S^{3}$, it instead induces $M-p$ units of D$3$ charge
(and tension). This is what is effectively describing the
annihilation, since one interprets this as if $M$ D$3$'s
materialized out of the fluxes and of which $p$ annihilated with the
anti-D$3$ branes, giving rise to the $M-p$ D$3$'s in the final
stage. The way charge conservation has been preserved throughout is
by the nucleation of a bubble of supersymmetric vacuum with one less
unit of $H$-flux (since one unit of $H$ flux induces $M$
D$3$-charges). The resulting picture of KPV is then that a
meta-stable state implies that the NS$5$-brane can find a local
minimum in energy by wrapping an $S^{2}$ of finite size and
therefore not move over to the other pole of the $S^{3}$
immediately.

It becomes then essential to understand whether the backreaction of
the anti-D$3$ branes changes this result in a significant way.
Therefore we first elaborate a bit on the validity of probe
approximations.

\subsection{The probe approximation}

Backreaction can invalidate a probe computation in two ways: 1) either the
backreacted solution is not localised in space or 2) it is not localised in time.

Planetary motion turns out to be a good illustration. When the mass $m$ of a
planet gets too large compared to the Sun's mass $M_{\bigodot}$, then the
planet is not moving in a geodesic of the Schwarzschild geometry formed by the
Sun. Instead, the sun will have an appreciable rotation around the common
centre of mass and the assumption of a Schwarzschild solution with fixed
position invalidates the probe approximation. This can be measured by how
delocalised the planet's geometry is compared to the Sun's geometry (i.e. is
there a large region in which the two Schwarzschild geometries have an
overlap). A good measure is the ratio $m/M_{\bigodot}$. Clearly, when
\begin{equation}
\frac{m}{M_{\bigodot}}\ll1\,,\label{inequality}%
\end{equation}
the probe computation works perfect and when
\begin{equation}
\frac{m}{M_{\bigodot}}\sim1\,,
\end{equation}
it fails, since the backreaction is not localised enough in space.

However, there are occasions in which inequalities such as
(\ref{inequality}) are fulfilled and still the probe approximation
fails. This happens exactly when the backreaction is not localised
in time. To see this, let us stick to a solar-like system, but this
time with several planets.  We insert an extra planet, or asteroid,
with a small mass that we hope to treat as just a probe. 
Nevertheless, the probe will eventually affect the motion of the
rest of the solar system in a non-negligible way. The better
$m<<M_{\bigodot}$ is satisfied, the longer it takes to see an
appreciable displacement in the positions of the other planets due
to the influence of the probe. If the system is chaotic, which
easily could be the case with several planets involved,  the
breakdown of the probe approximation will be quite dramatic provided
one waits long enough. To improve the situation, one can, after some
time, adjust the positions of the other planets and use these as new
inputs in the probe computation, and then continue the computation
until again the result deviates too much and so on.

The situation with the charged fluid is somewhat similar. The motion
of the negatively charged particle that we have inserted into the
fluid, can for some period of time be thought of as that of a probe
moving in the initial configuration of the fluid. Eventually, the
accumulated backreaction of the probe on the fluid will be too large
for this to work. At that point the new configuration of the fluid
needs to be used in order to correctly describe the subsequent
motion of the fluid and the probe.

The question then remains how to compute the corrections to the probe action due to the backreaction of the probe. Obviously one cannot fill in the backreacted solution into the probe action. Probe actions should be used such that one always subtracts the self energy of the probe. For instance the coupling of a charged particle, described by the source $j_{\mu}$, to the vector potential should really read
\begin{equation}
S_{probe} = \int j_{\mu}(A^{\mu}-A^{\mu}_{self})
\end{equation}
where $A^{\mu}_{self}$ is the vector potential that is locally generated by the charged particle. We already explained briefly how this works for a planetary system. If one would fill in the backreaction of the probe planet, the probe action would then describe geodesic curves that correspond to rotations around the Schwarzschild geometry of the planet instead that of the sun. One way to take into account corrections due to the probe backreaction is to realise the sun is rotating around a common center of mass. If one then would compute a probe action in a time-dependent geometry which describes a center that moves on a closed orbit, one would obtain a correction to the original geodesic curve. For the anti-branes the similar story would be that, to a first approximation, one only takes into account how the 3-form fluxes change due to the backreaction of the anti-brane. In other words, one has to fill in information about how the flux clumps, into the probe action. Whereas for the metric and 5-form field strength, one should take it the same as for the unbackreacted case. 

We will describe this in the next sections. But before we do, we want to emphasize that the intuition is rather clear. 
The basic picture is that the anti-brane will annihilate if it `feels' \emph{locally} too much D3 charge dissolved in fluxes. From the point of view of the anti-brane, it is irrelevant what happens far way.  This is similar to the positron that will annihilate with the electron cloud if the electrons get too close.  The backreaction of the anti-brane is exactly to increase the amount of D3 charge dissolved in fluxes, around the anti-brane. This will change its probability to annihilate from what is computed in the unbackreacted case. Therefore, prior to any computation, it should be obvious that the correction to the unbackreacted case is to make annihilation more likely. Hence such systems provide examples of where backreaction can significantly change the result although the backreaction might be very confined in space. The usual lore is that the probe action is valid if the  backreaction is not extending too far in space. But this is clearly incorrect for systems where the probe action only cares about what happens exactly near the probe.

\subsection{The flux-clumping parameter}

From the Bianchi identity (\ref{Bianchi}) we read off that the fluxes
$H$ and $F_{6-p}$ source a D$p$-brane charge density, $\rho$, of the
form
\begin{equation}
\rho\,\, \epsilon_{9-p}=H\wedge F_{6-p}\,,
\end{equation}
where $\epsilon_{9-p}$ are the directions transverse to the (anti-)Dp brane.
For the BPS backgrounds these fluxes obey the following relation
\cite{Blaback:2010sj}
\begin{equation}
H = \pm(g_{s})^{\tfrac{p+1}{4}} \star_{9-p} F_{6-p}\,,
\end{equation}
where the $\pm$ sign is a convention related to what one calls brane
and anti-brane charge.

When an anti-D$p$ brane is inserted into the BPS background the full
backreacted solution will have a complicated leg structure as can be
seen from the known solutions \cite{Bena:2009xk,
Massai:2011vi,Giecold:2011gw}. One exception is the backreaction for
the anti-D$6$, that can be captured by the following generalisation
of the BPS relation \cite{Blaback:2011nz}:
\begin{equation}\label{non-BPSflux}
H= \lambda(g_{s})^{\tfrac{7}{4}} \star_{3} F_{0}\,,
\end{equation}
where $\lambda$ is a function of the directions transversal to the
brane. This simplicity is the main reason that the anti-brane
backreaction is computable, near the anti-D$6$ branes, at all orders
in perturbation theory \cite{Blaback:2011pn}. One can T-dualise down
the non-compact anti-D$6$ solution, by taking some of the
worldvolume directions to be circles.  This provides solutions for
backreacted anti-D$p$ branes with $p<6$\footnote{We are grateful to
Iosif Bena for suggestions along this line.}. For the example of the
anti-D$3$ brane, this would give the backreacted solution on the
non-compact ``Calabi-Yau'' $R^3\times T^3$. However, close to the
anti-D$3$ brane this solution should be similar to the anti-D$3$
solution at the tip of the conifold. The T-dual of the non-BPS flux
relation (\ref{non-BPSflux}) becomes:
\begin{equation} \label{Ansatz}
H= \lambda(g_{s})^{\tfrac{p+1}{4}} \star_{9-p} F_{6-p}\,.
\end{equation}
From here on we refer to the function  $\lambda$  as the
\textit{flux-clumping parameter} as it reflects the relative amount
of flux that has been gathered near the anti-brane (compared with
the unperturbed BPS background). In what follows we will use the
Ansatz (\ref{Ansatz}) to revise the original KPV computation.
Another argument for why this simplified flux relation captures the
essential physics is simply that it is capable of describing the
clumping of the flux and one should regard the relation
(\ref{Ansatz}) as only being true in an integrated sense. For that,
consider the case of anti-D$3$ branes in the KS throat. The
$F_3$-flux is constrained to fulfill
\begin{equation}\label{eq:M}
M = \int_{S^3}F_3\,.
\end{equation}
The flux number $M$ provides us with a normalisation to measure the
clumping of the $H$-flux near the anti-D$3$ as follows
\begin{equation}\label{eq:lambda}
\lambda = -\frac{g_s^{-1}}{M}\int_{S^3}\star_6 H\,.
\end{equation}
This $\lambda$ then has to be compared with the one in
(\ref{Ansatz}).

As we have mentioned throughout the paper, the flux-clumping
parameter diverges near the anti-branes, for all known solutions.

\subsection{The KPV potential reconsidered}

In terms of the above analogy with planetary systems, the role of
the Sun and the larger planets in KPV is played by the charge
induced by the background flux in terms of the variable $\lambda$.
In fact, the analogy gets closer than we expect since brane-flux
annihilation can be understood in terms of the \textquotedblleft
geodesic\textquotedblright\ motion of a puffed up NS$5$-brane.

\begin{figure}[h] \centering
\includegraphics[width=.6\textwidth]{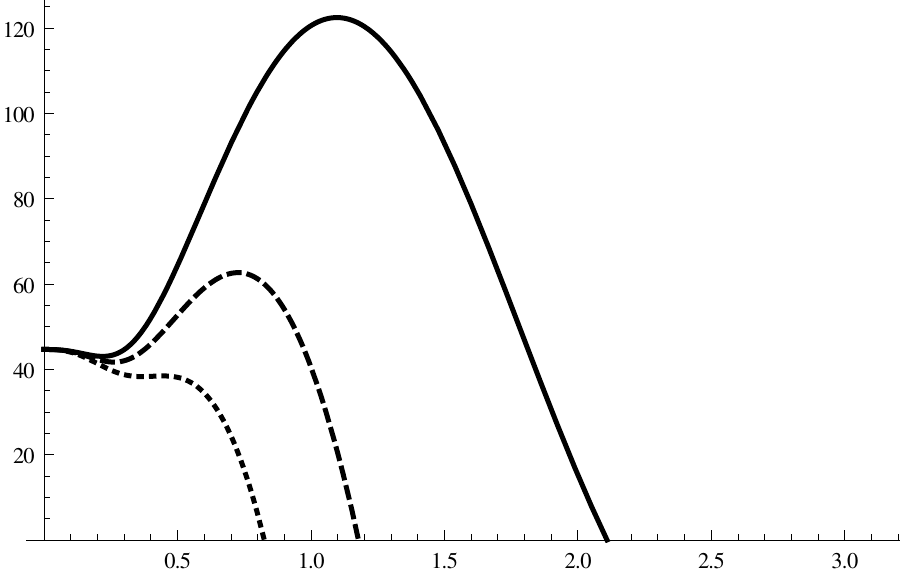}\caption{\emph{The
effective potential relevant for the NS$5$-motion, plotted for
different
values of $\lambda$.} }%
\label{Figure2}%
\end{figure}

If one assumes that the backreaction of $p$ anti-D$3$ branes is
confined in space, and can be made as small enough as we want by
tuning $p$ small, then the probe approximation can only fail when
the backreaction is not confined in time. This happens when the
\textquotedblleft position of the large planets\textquotedblright,
i.e. $\lambda$, changes in time. However small the effect of the
probe is, it will eventually cause the charged fluid to clump and
start to fall towards the probe. Near the probe itself the density
will diverge. For that reason we are interested in tracing back the
effect of $\lambda$ in the KPV potential, which enters through the
expression for $B_{6}$ that can be derived with the help of
(\ref{Ansatz}):
\begin{equation}
B_{6}\equiv\frac{1}{g_{s}^{2}}\star_{10}H=-\frac{\lambda}{g_{s}}V_{4}\wedge
F_{3}\,,
\end{equation}
where $g_{s}$ is the string coupling and $V_{4}$ is the red shifted
volume-form, along the four non-compact dimensions. Note again, as mentioned in discussion around (\ref{eq:M}) and (\ref{eq:lambda}), that this is only true in an integrated sense. Since $F_{3}$ is
topologically ``protected'' to have $M$ units of flux around the
$S^{3}$, we preserve the expression for $C_{2}$ from the KS
solution, such that \cite{Kachru:2002gs}
\begin{equation}
\int_{S^{2}}C_{2} = 4 \pi M (\psi-\tfrac{1}{2}\sin(2\psi))\,,
\end{equation}
where $\psi$ is the third Euler angle of the $S^{3}$, which measures
the sizes of the various $S^{2}$'s within $S^{3}$. Since the
NS$5$-brane wraps these $S^{2}$'s and moves along on the $S^{3}$,
$\psi$ is used to keep track of the position of the NS$5$-brane. If
the NS$5$-brane moves all the way to the other pole, where
$\psi=\pi$ it induces $M-p$ D$3$-charges instead of $p$ anti-D$3$
charges. The fact that the NS$5$-brane induces, initially, $p$
anti-D$3$ charge is due to the monopole charge of the worldvolume
flux $F_{2}$ of the NS5 brane, along the $S^{2}$
\begin{equation}
2\pi\int_{S^{2}} F_{2} = 4\pi^{2} p\,.
\end{equation}
With all these ingredients one can then compute the total
NS$5$-brane action as an effective action for the variable
$\psi(t)$, which captures the dynamics of the NS$5$-brane motion.
The effective potential one obtains is computed by putting the
momentum to zero in the higher-derivative Hamiltonian
\cite{Kachru:2002gs}, and it reads
\begin{equation}
\label{potential}\frac{V_{eff} (\psi)}{A_{0}} =\frac{1}{\pi} \sqrt{b_{0}%
^{4}\sin^{4}\psi+ \bigl(\pi\frac{p}{M}-\psi+\tfrac{1}{2}\sin(2\psi)
\bigr)^{2}} - \frac{\lambda}{2\pi}(2\psi-\sin(2\psi))\,,
\end{equation}
where $A_{0}$ is a constant of no relevance to our discussion and
the number $b_{0}\approx0.93266$, relates to the size of the apex in
the Klebanov--Strassler throat\footnote{The gravitational
backreaction of the flux-clumping is not related to the self-energy
of the anti-branes and hence needs to be taken into account. This
can qualitatively be done by changing the value of $b_0$. The change
is such as to only enhance the tendency towards an instability, just
like a large $\lambda$.}. The appearance of $\lambda$ in the
potential is very straightforward and to understand its impact we
show three plots in figure \ref{Figure2}.
The plot for $\lambda=1$ (full line), reproduces the meta-stable
vacuum of \cite{Kachru:2002gs}, with $p/M$ chosen as $3\%$. The dashed line
corresponds to $\lambda\approx1.3$ , in which the vacuum is barely there since
the barrier is lowered and the third plot, with dotted line, for which
$\lambda\approx1.7$ shows no meta-stable vacuum any more. As pointed out in
\cite{Kachru:2002gs} lower values of $p/M$ increase the height of the barrier.
To ensure that the stable minima will disappear for very small values of
$p/M$, $\lambda$ needs to be of the order $(p/M)^{-1/2}$.
This can be verified using a scaling such that $p/M \sim \epsilon^2$,
$\lambda \sim 1/\epsilon$, and $\psi \sim \epsilon$, where $\epsilon
\rightarrow 0$. In this limit the effective potential becomes
\begin{equation}
\frac{V_{eff} (\psi)}{A_{0}} =\frac{1}{\pi} \sqrt{b_{0}^{4}\psi^4+
\bigl(\pi\frac{p}{M}\bigr)^{2}} - \frac{2\lambda}{3\pi}\psi^3
\end{equation}
It can be easily checked that this potential lacks a metastable minimum if
$\lambda > \tfrac{b_0^6}{\pi}(p/M)^{-1/2}$.

\subsection{Interpretation}

\label{sec:interpretation}

It is not difficult to understand qualitatively why the barrier
vanishes for $\lambda$ greater than some critical value
$\lambda_{c}$ (for fixed $p/M$). The NS$5$-brane feels a force, due
to the fluxes, that drags it to other side of the $S^{3}$. However,
to move towards the other side it has to grow in size which is
energetically not favourable. Hence these two effects are played off
against each other and for the case backreaction is ignored
($\lambda=1$) there is a meta-stable state when $p/M$ is small
enough \cite{Kachru:2002gs}. For larger but constant $\lambda$ one
can get again a meta-stable vacuum, by lowering $p/M$ even more.

Let us for now assume that the backreacted anti-D$3$ solution in
the KS-throat displays the similar behaviour as found for anti-D$6$
branes in the supergravity analysis of \cite{Blaback:2011nz,
Blaback:2011pn} (and \cite{Blaback:2010sj}). Then this means that
$\lambda$ goes off to infinity near the anti-brane, which we
interpret as the non-existence of a static solution since $\lambda$
will grow in time as the flux cloud clumps. Within this
interpretation we have that, regardless of how small $p/M$, there
will be a point in time in which it has grown as to make the barrier
vanish and at that point the anti-branes annihilate with the fluxes.

This presents us with a conceptually complete and satisfying
picture. First, we have the supergravity equations that describe an
increasing clumping flux cloud. Second, open string effects kick in
and annihilate (a part of) the anti-branes. As a result we have one
smooth string theory solution in which clumping flux gradually
vanishes, leaving us finally with the SUSY ground state. This is a
consistent resolution of the singular flux found in the static
solution. This singular flux just came about because the real
process is time-dependent and the inability of the supergravity
degrees of freedom alone to produce the brane-flux annihilation,
results in an unphysical, infinite clumping of the flux in order to
create a static solution. Other examples of gravitating systems with
unusual singularities, that get resolved by time-dependence are
known to exist \cite{Gregory:1996dd, Dvali:2002pe}.

Although we presented the open string potential for the cases
applicable to anti-D$3$ branes in KS, one expects the same
qualitative results to hold for general D$p$-branes in backgrounds
with flux of the opposite charge. In particular for the case of
anti-D$6$ branes, for which the singular flux behaviour has been
established firmly \cite{Blaback:2011nz, Blaback:2011pn}. In the
latter case it is natural to expect that the system could puff up
into a fuzzy D$8$ that wraps the $S^{2}$-part of the transverse
geometry. 

\section{Conclusion}

In this note we give a possible interpretation of the flux singularities observed
for anti-brane solutions in warped throats  as
being due to the fatal attraction of the background fluxes towards
the anti-branes. When one insists on finding such solutions within a
stationary supergravity Ansatz one is bound to find singularities
that describe the infinite clumping of the fluxes. Hence, these
infinities are not an artefact of perturbation theory as suggested
in \cite{Dymarsky:2011pm}. This was already explicitly verified for
the case of anti-D$6$ branes \cite{Blaback:2011nz, Blaback:2011pn} 
and more recently for smeared anti-D3 branes \cite{Bena:2012bk}.

Secondly we have demonstrated that, once brane-flux annihilation is taken into
account, one obtains a sensible string theory solution in which fluxes are
drawn towards the anti-brane and subsequently annihilate. This gives rise to a
smooth time-dependent solution that approaches the supersymmetric vacuum at
large times. This we have verified by demonstrating that the effect of
accumulating fluxes is to lower the barrier against perturbative brane-flux
annihilation. This means that for a given $p/M$ there exists a critical flux
density, given by $\lambda_{c}$, such that for $\lambda>\lambda_{c}$ the
barrier vanishes. Since the real solution is time-dependent with an
ever-growing $\lambda$, as we argued, the anti-branes will annihilate and
there is no meta-stable state \footnote{After the first version of this paper came out, reference \cite{Kutasov:2012rv} came out which also discusses the KPV scenario from a type IIA point of view. The findings of \cite{Kutasov:2012rv} our different from KPV and from our conclusions. It would therefore be interesting to understand our interpretation from a IIA point of view to shed light on the different interpretations. }.

There is one possible caveat to our interpretation of the singular
fluxes, which was mentioned in \cite{Dymarsky:2011pm}. Close to the
anti-D$p$ branes, in the infrared region, the real solution can be different from a 
localised anti-brane and instead be a fuzzy D$(p+2)$-brane wrapping an $S^{2}$.
This means that the real supergravity solution will get altered near
the anti-branes. Given our simple analogy with negative particle inserted in
a positively charged cloud\footnote{If one would try to make a physical, non-imagined, setup like this, one could drop a positron in an electron cloud. When the electrons come closer to the positron they can first form positronium, which can indeed be a metastable state, although extremely short lived. The reason a meta-stable state is possible comes from the quantisation of the energy levels for the orbits. The analogues statement with anti-branes would exactly be the barrier in the KPV potential, which we argued is not there.}, the analogous statement would be that the particle
ceases to be a point particle but effectively gets smeared out a
bit. One can then imagine that the flux clumping will be less
severe, depending on how much the particle is smeared out\footnote{ However,
a counter-example of this appeared in \cite{Blaback:2011nz} in which
the anti-D$6$ brane delta-function support was replaced by an
arbitrary smooth profile. The computations revealed that there is
only a stationary supergravity solution in the case the profile is a
constant everywhere.}.  This has been 
investigated for the anti-D6 solution in \cite{Bena:2012tx}, where it was found that the singularity will indeed disappear if the anti-D6 would polarise into a spherical D8 brane.  However it was shown in detail that there is no polarisation taking place \cite{Bena:2012tx}  (although all the ingredients for polarisation are present in the solution).

If a similar statement is true for the anti-D3 brane (where it would be the D$5$ polarisation channel that could resolve the singularities in the 3-form fluxes), it implies we have to revise the picture of a de Sitter landscape
in string theory.

If, on the other hand, the singularities would get resolved by brane
polarization, it still remains to be demonstrated that a meta-stable
state exists, because it could easily be that the flux-clumping,
although finite, exceeds the critical value $\lambda_{c}$, such that
the system is perturbatively unstable against anti-brane
annihilation. If, after brane polarization, $\lambda$ is indeed
finite and smaller than the critical value $\lambda<\lambda_c$ its
size will still be larger than in the unbackreacted case
($\lambda>1$) and the bounds on $p/M$ will be tighter, something the
authors of \cite{Kachru:2002gs} remarked themselves, though without
any argument.

It is therefore a very interesting problem to find (properties of)
the fully backreacted solution in which the anti-branes could be
polarised.
%

\section*{Acknowledgments}

We benefitted from useful discussions with Iosif Bena, Gregory
Giecold, Stanislav Kuperstein, Stefano Massai and Bert Vercnocke. JB
is supported by the G{\"o}ran Gustafsson Foundation. UD is supported
by the Swedish Research Council (VR) and the G\"oran Gustafsson
Foundation. TVR is supported by the ERC Starting Independent
Researcher Grant 259133-ObservableString.

\bibliographystyle{utphysmodb}
\bibliography{refs}

\end{document}